# Mueller matrix maps of dichroic filters reveal polarization aberrations

James Heath[a], Meredith Kupinski[a], Ewan Douglas[b], Kira Hart[a], and James Breckinridge[a,c]

[a]Wyant College of Optical Science, University of Arizona, Tucson, USA
[b]Steward Observatory, University of Arizona, Tucson, USA
[c]Astronomy, California Institute of Technology, Pasadena, USA

## ABSTRACT

Dichroic filters are used by instrument designers to split a field of view into different optical paths for simultaneous measurement of different spectral bands. Quantifying the polarization aberrations of a dichroic is relevant for predicting the incident polarization states downstream, which could affect the performance of diffraction limited systems. One important application is the fore-optics of exoplanet imaging coronagraphs. In this work, the polarization properties of the Edmund #69-205 650 nm roll-off dichroic are measured using a rotating retarder Mueller matrix imaging polarimeter. The polarization properties of this commercial dichroic are compared at normal and 45° angle of incidence. The normal incidence measurements verify the instrument calibration since no polarization aberrations were observed. Transmission measurements at 680 nm and 45° yield a 2.9 rad magnitude of retardance and 0.95 diattenuation. Effectively, at 630 nm the dichroic is a $\lambda/4$ waveplate with a horizontal fast-axis.

## 1. INTRODUCTION

High contrast imaging of exoplanet systems is challenging because the planet is many orders of magnitude dimmer than the neighboring star. Vector vortex coronagraphs (VVCs) utilize a series of polarization elements to discard the light from the star and allow exoplanet imaging. The achievable contrast of these systems is limited by unwanted polarization artifacts generated as light propagates through the system.[1] Many VVC designs utilize dichroic filters to remove light from the optical path outside the wavelength specifications of subsequent optics or the detector. An ideal dichroic filter, as often assumed in VVC optical models, would not contribute any retardance nor diattenuation to the system. However, this assumption is not valid. Since dichroic bandpass filters are constructed by depositing thin films across the face of a dielectric substrate, the Fresnel equations dictate that varying the angle of incidence (AOI) from the normal will introduce some degree of retardance and diattenuation. The simplest way to reduce this type of error in general is to design the system such that rays intersecting each surface have the lowest AOI possible. However, as dichroics are designed to operate at AOI = 45° to achieve high contrast after the roll-off wavelength, this is not feasible. Another common mitigation strategy to reduce this type of polarization error is to utilize subsequent optics to compensate for the introduced retardance and diattenuation. For example, some setups will use two fold mirrors, the second to compensate the polarization aberrations of the first by aligning their s- and p- polarization states between them.[2] By measuring the magnitude and variation of diattenuation and retardance across the dichroic using Mueller matrix polarimetry, the feasibility and appropriateness of using compensating optics to correct for any introduced error can be quantified.

Breckinridge et al. have reported that polarization aberrations are introduced into systems when: 1) rays intersect surfaces at non-normal incidence, 2) when reflections occur from metals, 3) when wavefronts pass through dielectric substances like windows and glass filters, and 4) systems using diffraction gratings and prisms.[1] When operating at the intended AOI, dichroic filters are usually modeled as a mirror at wavelengths longer than

Further author information: (Send correspondence to James C. Heath)
James C. Heath: E-mail: heathjam@email.arizona.edu

the roll-off wavelength and as a transmitting plane parallel plate for shorter wavelengths. This is accurate for non-diffraction limited optical systems that are adequately modeled using geometrical optics. Polarization preserving, diffraction limited systems for scientific investigations require vector-wave modeling tools to describe performance. Dichroics satisfy potentially three of the four sources of polarization aberrations that can be introduced to high-contrast systems. Davis et al. have revealed that the process of depositing these thin film coatings can leave amorphous microscopic grain structures that can exhibit spatially-varying retardance.[3] Since dichroics are fabricated using thin film deposition, investigating the magnitude of spatial dependence from these optical elements is paramount to ensure high-contrast performance in next-generation systems.

Several space telescopes and instruments used for astronomy and Earth science utilize dichroic filters for optical wavefront division. This enables simultaneous imaging of incoherent white-light at multiple wavelengths. Dichroic filters are used in the NASA HabEx, Roman (WFIRST), and LUVOIR optical systems. The efficiency of the dichroic's interference is wavelength-dependent due to wavelength-dependent polarization effects. Breckinridge, et.al.[4–6] show that image formation is a process of interferometry to show that image quality depends on how optical polarization changes across the surface of wavefronts and is most noticeable in diffraction limited optical systems. Optical system image quality and system transmittance/reflectance depends on how the polarization properties of dichroics change with wavelength. Most optical designers and engineers along with astronomers and Earth scientists currently assume that dichroic filters have no effect on image quality. Here we report Mueller matrix measurements of a commercial dichroic to inform the design, specification, and simulation of these imaging systems.

# 2. METHODS

## 2.1 Mueller Matrix Polarimetry

A Mueller matrix polarimeter is used to measure the diattenuation, depolarization, and retardance of a commercial dichroic. The polarization state generator (PSG) creates a set of $q$ polarization states, denoted by the Stokes vector $\mathbf{S}_q$, as a retarder is rotated. This light is transmitted through the sample and is modelled by the Mueller matrix-Stokes relation $\mathbf{M} \cdot \mathbf{S}_q$ where $\mathbf{M}$ is the Mueller matrix of the sample. The $q$th exiting Stokes parameters are then analyzed by a polarization state analyzer (PSA) and the polarization-dependent transmission is given by the Stokes vector $\mathbf{A}_q$. The $q$th measured flux is $P_q = \mathbf{A}_q^T \mathbf{M} \mathbf{S}_q$. This linear expression relates the measured flux to the sample's Mueller matrix elements. From multiple measurements at varying retarder rotations a set of linear equations can be inverted to solve for the sample's Mueller matrix.[7]

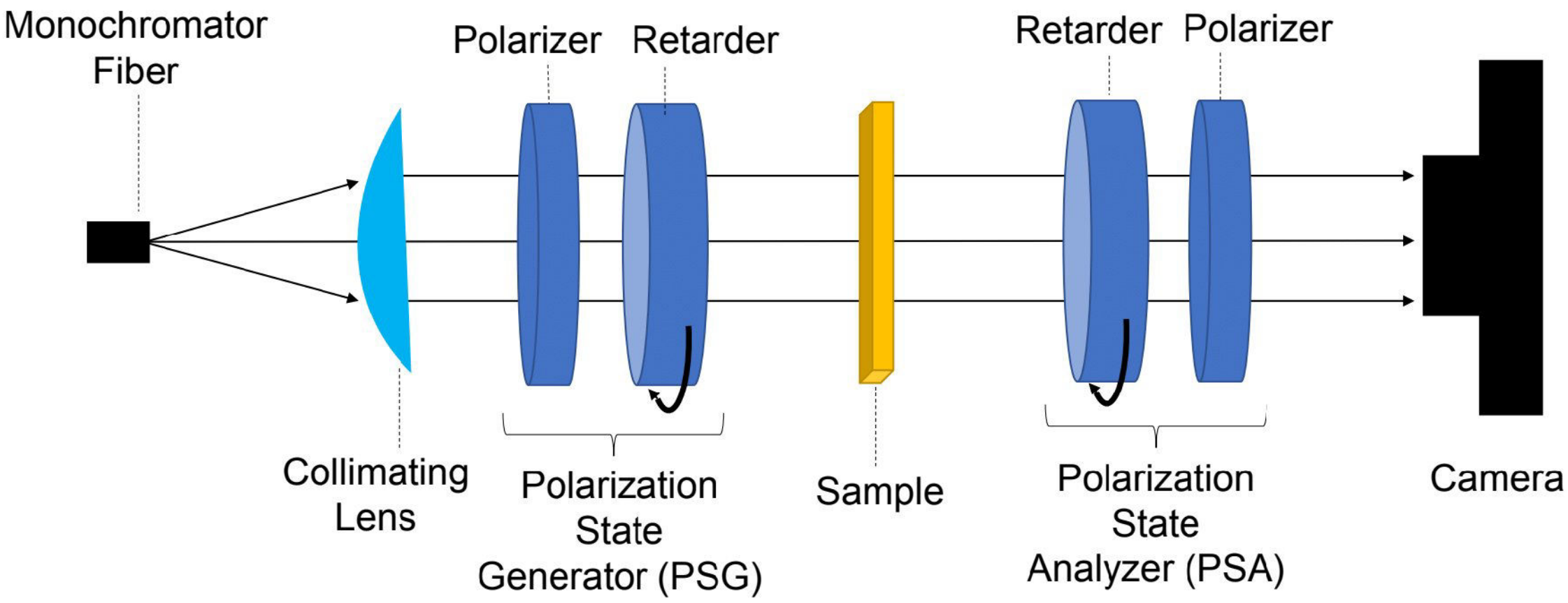

Figure 1: Schematic of the Mueller matrix stepper polarimeter. The illumination system is comprised of a fiber source and collimating optics. A dual rotating retarder creates polarization modulation in the system. A camera with a CMOS detector is conjugate to the sample.

The Mueller matrix stepper polarimeter consists of four major subsystem interfaces. A specular illumination system is comprised of a fiber coupled to a xenon light source and collimating lens. A dual rotating retarder polarization modulation system with linear polarizers provide polarization control as the acting PSG and PSA. This leads into an imaging lens to focus onto a digital CMOS camera. Figure 1 includes the arrangement and labels the four main subsystems.

The illumination optics consist of an Axometrics Tunable Visible Source (TVS) which provides monochromatic light in the range of 400-800 nm into a 200 mm focal length collimating lens. The light source is a 150 W Xenon arc lamp that ranges from 400-800 nm. The TVS has a wavelength accuracy of $\pm$ 0.5 nm at 550 nm and 4 nm bandwidth across the full range.

The PSG and PSA consist of two rotating mounts holding custom achromatic retarders that operate at $\approx \lambda/3$ and two fixed linear polarizers. The PSG varies the incident polarization state of the light into the sample, while the PSA controls the exiting polarization state into the camera. This allows for intensity measurements at various polarization configurations to provide the necessary Stokes information for analysis.

The imaging optics consist of an Edmund Optics focusing lens directly attached via c-mount to a Hamamatsu C11440-42U30 ORCA-Flash4.0 LT+ Digital CMOS Camera. The camera sensor has 2048 x 2048 square pixels, each 6.5 $\mu$m x 6.5 $\mu$m and have a bit depth of 16-bits. It provides 4.0 megapixel resolution at 30 fps while achieving 0.9 electrons (median) 1.5 electrons (r.m.s) readout noise performance and achieves reasonable levels of quantum efficiency in the measurement range. Measurements are taken across a 1500 x 1500 region of interest (ROI) to improve computation times.

The dichroic (catalog #69-205) measured was an Edmund Optics 25 mm diameter fused silica low pass dichroic filter with a roll-off at 650 nm and built to be used at 45° with 97% average polarization reflection and 80% transmission. Ideally, the Mueller matrix of the dichroic would be the identity matrix, not introducing any diattenuation, retardance, or depolarization into the system.[8]

In this experiment a series of polarimetric measurements are taken over 64 different orientations of the PSG and PSA to reconstruct the Mueller matrix of the dichroic. Two full measurements of the dichroic are taken at 0° and 45° AOI for validation. The 0° measurement served as a control because it is not expected to measure high values of diattenuation or retardance at an normal incidence. An ideal dichroic would have only polarization independent transmission with no diattenuation, retardance, or depolarization, and have little to no spatial variation. However, real dichroics must have diattenuation for non-zero AOI as described by the Fresnel equations. It is expected that for any thin film element there will be some amount of retardance at varying AOI from the normal. Along with this, scattering from the element will generate some degree of depolarization. The question is whether these effects will make a large impact on the performance of high contrast systems such as vector vortex coronagraphs.

### 2.2 Mathematical Methods

Mueller calculus is used to describe the polarization state of light and the effect materials have on the polarization state of light. The four Stokes parameters describe all possible polarization states of light and are defined as

$$\mathbf{S} = \begin{bmatrix} S_0 \\ S_1 \\ S_2 \\ S_3 \end{bmatrix} = \begin{bmatrix} P_H + P_V \\ P_H - P_V \\ P_{45} - P_{135} \\ P_R - P_L \end{bmatrix}, \quad (1)$$

where $\mathbf{S}$ is called the Stokes vector and P are irradiance measurements in units of [W/m$^2$]. A capital $\mathbf{S}$ indicates the Stokes vector is unnormalized. The subscripts on P denote transmission through a polarization filter: horizontal linear (H), vertical linear (V), 45° linear, 135° linear, right-circular (R), and left-circular (L). The same propagating linearly polarized light measured from two different rotated detector positions may appear to return different orientations.[9]

The 4 $\times$ 4 Mueller matrix M is a representation of a materials' linear interaction with the Stokes parameters

$$\mathbf{M} \cdot \mathbf{S} = \mathbf{S}' = \begin{bmatrix} M_{00} & M_{01} & M_{02} & M_{03} \\ M_{10} & M_{11} & M_{12} & M_{13} \\ M_{20} & M_{21} & M_{22} & M_{23} \\ M_{30} & M_{31} & M_{32} & M_{33} \end{bmatrix} \begin{bmatrix} S_0 \\ S_1 \\ S_2 \\ S_3 \end{bmatrix} = \begin{bmatrix} S'_0 \\ S'_1 \\ S'_2 \\ S'_3 \end{bmatrix}, \quad (2)$$

where $\mathbf{S}$ and $\mathbf{S}'$ are the Stokes vectors for incident and exiting states respectively, necessarily making $\mathbf{M}$ a 4 x 4 matrix. The Mueller matrix provides a systematic way of representing all of the polarization properties of a sample: diattenuation, retardance, and depolarization. Since the elements of $\mathbf{S}$ and $\mathbf{S}'$ are irradiances, $\mathbf{M}$ are dimensionless ratios of said irradiances and are all real-valued. To help interpretation, the Mueller matrix is often normalized in reference to the $M_{00}$ element. A normalized Mueller matrix and its elements are denoted by lower-case $\mathbf{m}$ and $m_{ij} = M_{ij}/M_{00}$ respectively.

Once a Mueller matrix of the sample is measured, polarization characteristics such as diattenuation, retardance, and depolarization metrics can be extracted using the Lu-Chipman decomposition.[10] In this algorithm, the sample Mueller image M is decomposed pixelwise into a product of three matrices

$$\mathbf{M} = \mathbf{M}_{\Delta}\mathbf{M}_{R}\mathbf{M}_{D}, \tag{3}$$

where $\mathbf{M}_{\Delta}$, $\mathbf{M}_{R}$, and $\mathbf{M}_{D}$ are the matrices of a depolarizer, a retarder, and a diattenuator, respectively. The decomposition can be implemented for any Mueller matrix, and provides unique results. The scalar retardation and diattenuation as well as the corresponding eigenpolarizations of the sample under study are then defined as those of the component matrices $\mathbf{M}_{R}$ and $\mathbf{M}_{D}$.[11]

Diattenuation describes the polarization dependence of the transmittance and can be calculated directly from the Mueller matrix without decomposition by the following equation

$$D = \frac{T_{max} - T_{min}}{T_{max} + T_{min}} = \frac{\sqrt{M_{01}^2 + M_{02}^2 + M_{03}^2}}{M_{00}}, \tag{4}$$

where $T_{max}$ and $T_{min}$ are the measured maximum and minimum transmitted irradiances. The diattenuation is only dependent on the first row of the Mueller matrix because these are the only elements that affect $S_0'$. Therefore, the diattenuation can be written as a total magnitude of this first row, normalized by the overall transmittance, $M_{00}$, which constrains the diattenuation from $0 \leq D \leq 1$.[7] Retardance is described as the optical path or phase difference between two polarization states that may simply be described as $\delta = \phi_s - \phi_p$, where $\phi_s$ and $\phi_p$ represent s- and p- polarized light respectively. Retardance can have a significant effect in polarization sensitive applications such as vector vortex coronagraphy.

A Jones matrix is a useful and widely used formalism for describing the polarization effects of an optical element on fully-polarized, monochromatic light. A Mueller matrix measurement can be converted to a Jones matrix if depolarization and absorption effects are neglected. A useful decomposition of a Mueller matrix is as a convex sum of four non-depolarizing Mueller matrices

$$\mathbf{M}_d = \lambda_1\widehat{\mathbf{M}}_\mathbf{1} + \lambda_2\widehat{\mathbf{M}}_\mathbf{2} + \lambda_3\widehat{\mathbf{M}}_\mathbf{3} + \lambda_4\widehat{\mathbf{M}}_\mathbf{4} \tag{5}$$

where the scalar values are rank-ordered such that $\lambda_1 > \lambda_2 > \lambda_3 > \lambda_4 > 0$ and $\widehat{\cdot}$ denotes a non-depolarizing Mueller matrix.[12,13] If only $\lambda_1$ is non-zero then $\mathbf{M}_d$ is non-depolarizing. To estimate the closest non-depolarizing Mueller matrix from a general depolarizing Mueller matrix, only the first term of Eq. 5 is retained. A non-depolarizing Mueller matrix can then be converted to a Jones matrix. Conversion from the Mueller matrix to a Jones matrix for the dichroic was performed using the Airy Optic's Polaris-M program, which has a built in function for interpolating depolarizing Mueller matrix images into a Jones matrix representation using this method.[14] Once converted to a Jones matrix, Mueller matrix measurements of a sample can then be used in Jones matrix ray-tracing models to quantify the sample's contributions to the Jones pupil of the full optical system.

## 3. RESULTS & DISCUSSION

### 3.1 Wavelength Dependence

Figure 2 shows the spatially averaged Mueller matrix measurements for a central 10×10 pixel region of the dichroic from 450-750 nm in 10 nm steps. The $M_{00}$ element, which is equivalent to total transmission, demonstrates higher transmission above the 650 nm roll-off when AOI=0° and at 45° the transmission agrees with manufacturer specifications. This result is expected and is used as a validation to show that as the AOI increases the pass band is blue shifted.

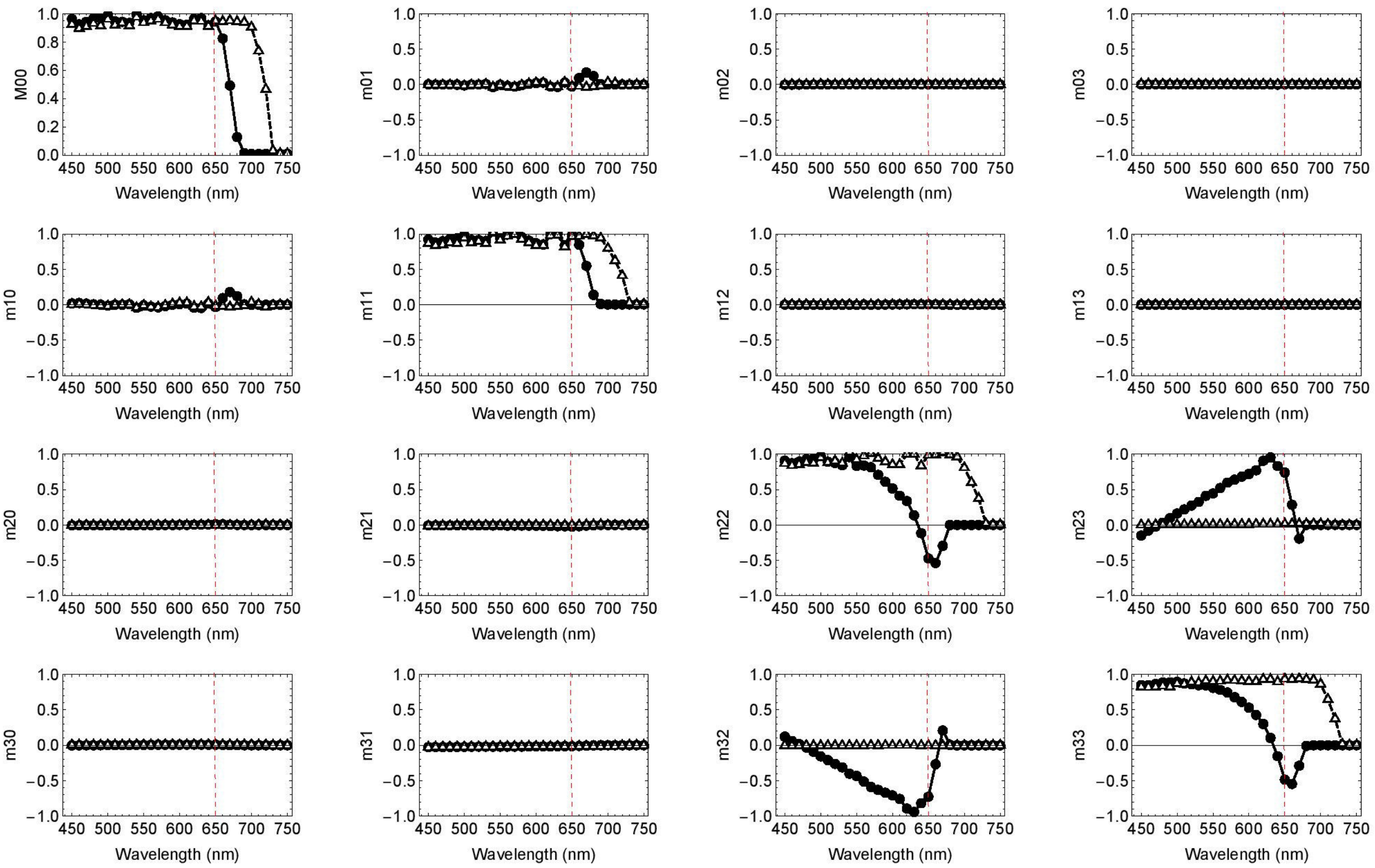

Figure 2: Mueller matrix measurements of the dichroic from 450-750 nm in 10 nm steps. Triangle markers are AOI=0° and circle markers are AOI=45°. The vertical red line is at the roll-off wavelength of 650 nm. Measurements are averaged over a central $10 \times 10$ pixel region. The 15 non-$M_{00}$ elements have been normalized by $M_{00}$ which is denoted with a lowercase $m_{ij}$.

The $m_{23}$ and $m_{32}$ elements in Figure 2 show coupling between $\mathbf{S}_2$ and $\mathbf{S}_3$ for AOI=45° which indicates retardance. This retardance is not measured at normal incidence. At 630 nm, the $m_{22}$ and $m_{33}$ elements are 0 and the dichroic response is nearly an ideal horizontal fast-axis $\lambda/4$ waveplate.

Figure 3a shows the diattenuation of the dichroic at 0° and 45° AOI. As expected from the Fresnel equations, the diattenuation is negligible at normal incidence, but at 45° AOI increases rapidly at above the roll-off wavelength. At 680 nm, diattenuation peaks at approximately 0.95 and then decreases back down to near zero at 700 nm. Non-negligible diattenuation starts at approximately 660 nm which indicates a difference in transmission between the s and p polarization states at wavelengths higher than the dichroic roll-off. This difference in transmission is identified by the manufacturer in their s and p polarization transmission graphs.[8]

In Figure 3b, the average retardance magnitude at 45° AOI shows a significant increase starting from 470 nm and reaching a $\pi/4$ shift at approximately 590 nm. This continues to grow to $\pi/2$ at 630 nm. As the transmission in the $m_{22}$ and $m_{33}$ elements decrease along with this retardance shift, the dichroic starts exhibiting quarter waveplate behavior when nearing 630 nm as seen Figure 2. This retardance increases through the roll-off, approaching halfwave plate behavior at 680 nm, before dropping rapidly at 690 nm and stagnating at $\pi/4$. The response in the Mueller matrix after 700 nm are noisy, making it difficult to draw any distinct conclusions from this portion of the curve, though it should be noted that the retardance stays at around $\pi/4$ in this range.

Figures 4a and 4b show the Stokes parameters exiting the dichoric as a function of wavelength for unpolarized and right circularly polarized (RCP) inputs. Unpolarized light remains largely unpolarized until after the roll-off. Looking at $S_1$ reveals that the polarizing properties due to diattenuation still exist in the optic for unpolarized

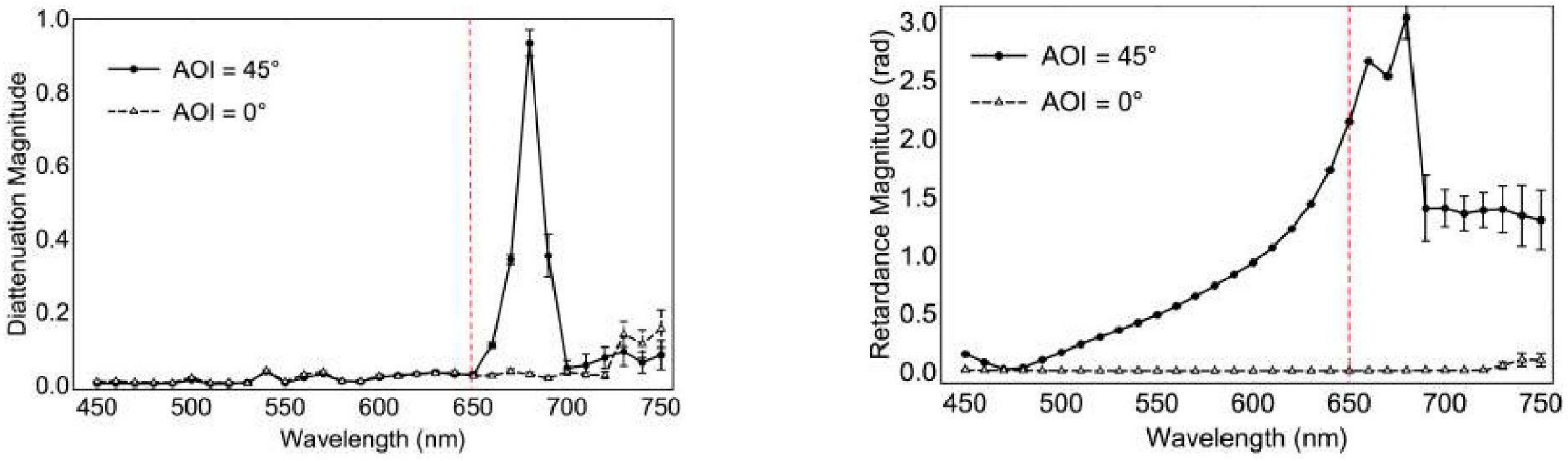

(a) Diattenuation

(b) Retardance Magnitude

Figure 3: In (a) the diattuation (see Eq. 4) and in (b) the retardance magnitude (see Eq. 3) of the Mueller matrix measurements from Figure 2. Triangle markers are AOI=0° and circle markers are AOI=45°. The retardance and diattenuation are negligible at normal incidence as expected. At 45° AOI, diattenuation is introduced at 660 nm, peaking near 0.95 at 680 nm and then falling off rapidly above 700 nm. Retardance is measured at 460 nm and increases to 2.25 rad at 650 nm and up to 2.9 rad at 680 nm, before falling off above 690 nm.

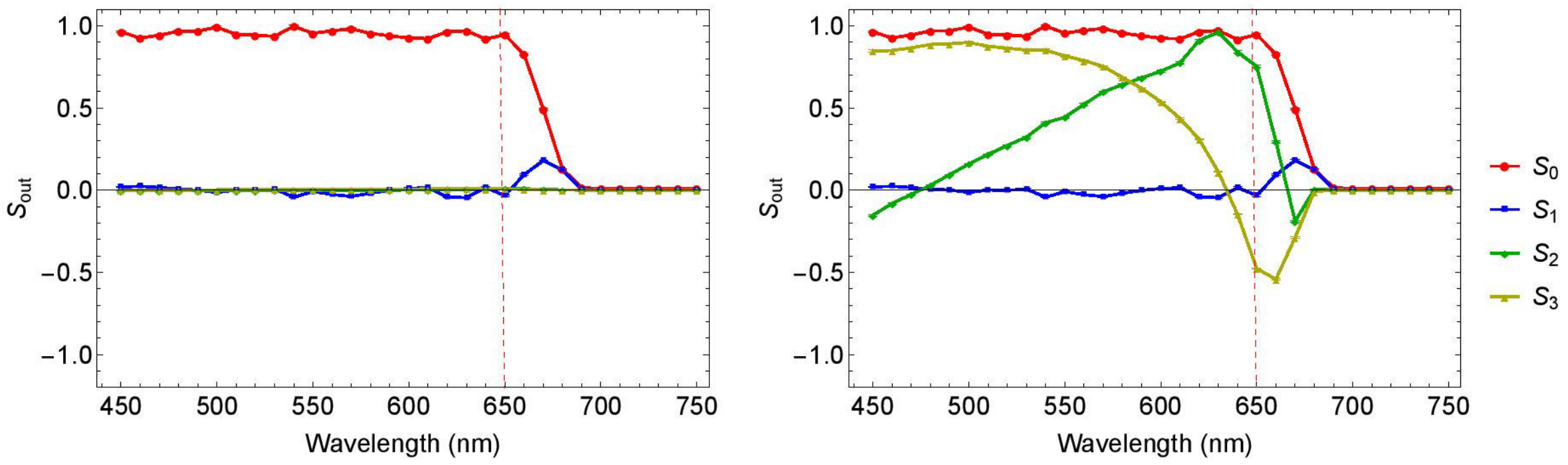

(a) Unpolarized $\mathbf{S}_{in}$

(b) Right Circular Polarized (RCP) $\mathbf{S}_{in}$

Figure 4: Simulated Stokes parameters of light transmitted through the dichroic as a function of wavelength. Using the relation $\mathbf{S}_{out} = \mathbf{M}\mathbf{S}_{in}$ for (a) $\mathbf{S}_{in} = (1, 0, 0, 0)$ and (b) $\mathbf{S}_{in} = (1, 0, 0, 1)$ each plots displays components of $\mathbf{S}_{out} = [S_0(\text{red}), S_1(\text{blue}), S_2(\text{green}), S_3(\text{yellow})]$. In (a) unpolarized input remains unpolarized below the roll-off wavelength. Above the roll-off, from 660-680 nm, diattenuation of the dichroic induces polarizance. In (b) the retardance of the dichroic rotates a RCP input at all wavelengths. The RCP input exits as 45° at 630 nm.

light right after the roll-off wavelenghth starting at 660 nm and peaking at 680 nm. As seen in Figure 4b, incoming RCP light's polarization state shifts significantly approaching the roll-off range. At approximately 635 nm, the incoming RCP light will exit in a linear state with a 45° orientation, which is evidence of the $\lambda/4$ waveplate like response of the Mueller matrix at 630 nm.

### 3.2 Spatial Dependence

Spatial dependencies of the dichroic at 45° are measured at 450 nm, 630 nm, 650 nm, and 670 nm. The mark on the right side of each of the following graphics is a fiducial that was used for alignment. Due to little spatial dependence for the on-diagonal amplitudes and negligible values in their off-diagonals, only the phase terms of the Jones plots are analyzed to investigate the retardance orientation. Table 1 summarizes the total range in diattenuation and retardance across the dichroic at different wavelengths. Spatial plots of diattenuation and retardance over the aforementioned wavelengths are shown in Figures 5 and Figure 6 respectively.

Figure 5 shows negligible values in spatial diattenuation until the roll-off wavelength. At 450 nm, there is little diattenuation, with a maximum around 0.03 in the lower right region and an overall difference of ~0.02 across the pupil showing slight tilt. There no increase in diattenuation at 630 nm, reaching a maximum value of 0.01 across the pupil, The diattenuation raises slightly at 650 nm to a value of 0.04 in the lower right hand of the pupil, but reaching almost 0.1 in the upper left of the pupil. At 670 nm, a significant diattenuation gradient is visible, at 0.9 in the upper left of the pupil and 0.81 in the lower right, making a 0.09 difference across the region. This diattenuation was found in the 10x10 pixel region during spectral analysis, however the tilt across the pupil shows that the entire region needs to be taken into account when observing these properties.

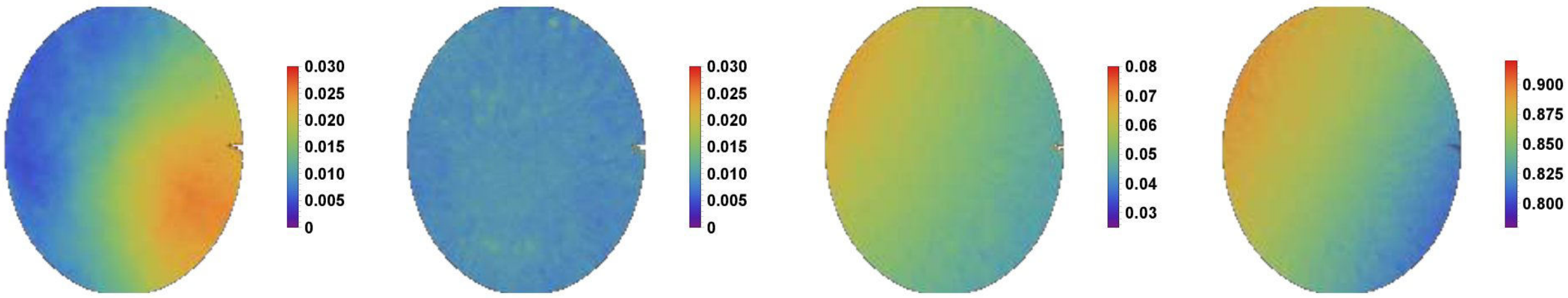

(a) 450 nm (b) 630 nm (c) 650 nm (d) 670 nm

Figure 5: Spatially-varying diattenuation of the dichroic at four selected wavelengths. The pupils appear elliptical due to 45° AOI of the dichroic. The overall diattenuation changes with wavelength in this range so each color bar is rescaled in (a-d). The diattenuation is smoothly varying over the optic and has a minimum to maximum range of approximately 0.025 in (a) 450 nm and 0.087 at (d) 670 nm. On the right edge of the optic the fiducial mark is visible.

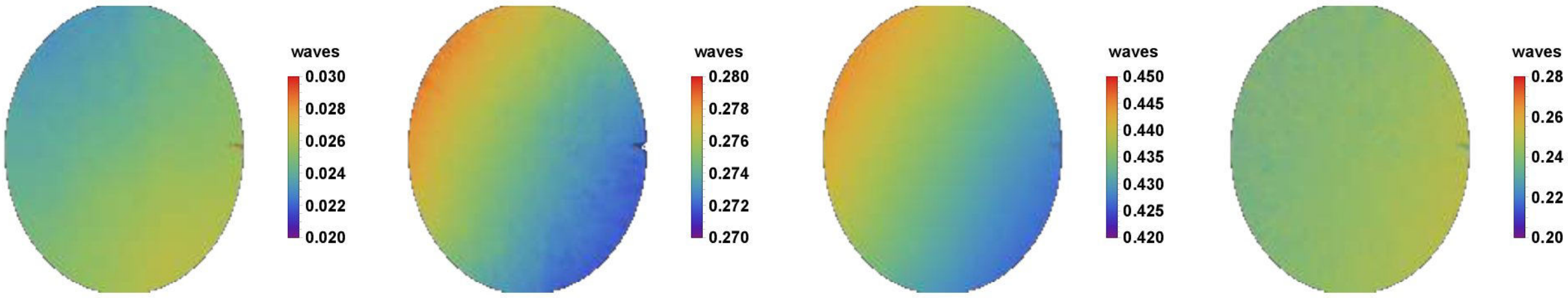

(a) 450 nm (b) 630 nm (c) 650 nm (d) 670 nm

Figure 6: Spatially-varying retardance magnitude at four selected wavelengths. The pupils appear elliptical due to 45° AOI of the dichroic. The retardance magnitude is a strong function of wavelength so each color bar is rescaled in (a-d). The retardance magnitude is smoothly varying over the optic and has a minimum to maximum range in (a) $\approx$ 0.01 waves, (b) $\approx$ 0.01 waves, (c) $\approx$ 0.03 waves, and (d) $\approx$ 0.06 waves. On the right edge of the optic a fiducial mark is visible.

| Wavelength | Diattenuation | Retardance [waves] |
|---|---|---|
| 450 nm | 0.005 - 0.030 | 0.024 - 0.027 |
| 630 nm | 0.010 - 0.013 | 0.272 - 0.279 |
| 650 nm | 0.042 - 0.076 | 0.426 - 0.446 |
| 670 nm | 0.800 - 0.887 | 0.239 - 0.258 |

Table 1: Range of measured diattenuation and retardance across the dichroic summarizing the spatially varying results shown in Figures 5 and 6.

| Wavelength | $\tau$ | Jones Matrix Phase [waves] |
|---|---|---|
| 450 nm | 0.98 | $\begin{bmatrix} 0.012^{+0.00}_{-0.01} & -0.12^{+0.48}_{-0.13} \\ -0.094^{+0.12}_{-0.15} & -0.01^{+0.01}_{-0.00} \end{bmatrix}$ |
| 630 nm | 0.94 | $\begin{bmatrix} -0.14^{+0.14}_{-0.00} & -0.41^{+0.42}_{-0.01} \\ 0.44^{+0.03}_{-0.44} & 0.14^{+0.00}_{-0.14} \end{bmatrix}$ |
| 650 nm | 0.89 | $\begin{bmatrix} -0.23^{+0.23}_{-0.00} & -0.37^{+0.72}_{-0.02} \\ -0.27^{+0.28}_{-0.21} & 0.23^{+0.00}_{-0.23} \end{bmatrix}$ |
| 670 nm | 0.31 | $\begin{bmatrix} 0.13^{+0.01}_{-0.13} & 0.34^{+0.05}_{-0.48} \\ 0.19^{+0.08}_{-0.36} & -0.13^{+0.13}_{-0.01} \end{bmatrix}$ |

Table 2: Measured and calibrated average Jones matrix pupil phase-maps across the dichroic are presented in the $xx$, $xy$, $yx$, $yy$ format at 450, 630, 650, and 670 nm in units of waves, summarizing the results in Figure 7. Matrix elements are given in the form $avg^{\Delta max}_{\Delta min}$, where $\Delta max$ and $\Delta min$ are the difference between the mean and extremum values found across the aperture of the dichroic, not including values near the fiduciary. The average transmittance of the dichroic ($\tau$) at these wavelengths is also given. Values of $\pm 0.00$ indicate that the extremum value is less than 1/100th of a wave from the mean.

In Figure 6, retardance across the pupil is relatively low at approximately 0.025 waves and a change of 0.01 waves across the region. At 630 nm there is an increase to a little over 0.27 waves of retardance, with a difference of 0.01 waves from upper left to the lower right of the pupil. This retardance gradient pattern continues at 650 nm, but with an increase of magnitude up to an approximate maximum of 0.45 waves of retardance and a difference across the pupil of $\approx 0.03$ waves. At 670 nm, the retardance magnitude falls to around 0.25 waves, with a 0.06 wave difference across the region.

Figure 7 shows Jones pupil phase-maps across the dichroic in waves at 450, 630, 650, and 670 nm and Table 2 gives the average value of each pupil and its transmittance ($\tau$). In the on-diagonal phase terms at 450 nm, there is <0.02 waves across the pupil's field. The $\phi_{xy}$ term shows a positive change in phase of approximately half a wave from upper left to the lower right of the pupil. At 630 nm, there is an increase in phase in the on-diagonals reaching average values of -0.14 waves in $\phi_{xx}$ and +0.14 waves in $\phi_{yy}$. The off diagonals are of opposite sign and equal in magnitude at approximately $\pm 0.4$ waves, with a difference of 0.85 waves between them. This difference in the off-diagonals reduces at 650 nm to 0.1 waves. The $\phi_{xx}$ and $\phi_{yy}$ terms at 650 nm show a slight increase in their magnitudes to a approximately $\pm 0.23$ waves. The on-diagonals at 670 nm shows a sign inversion and decrease in magnitude to $\pm 0.13$ waves. The off-diagonals also show a change in sign and reduction in magnitude to approximately 0.34 waves in the $\phi_{xy}$ term and 0.19 waves in the $\phi_{yx}$ term.

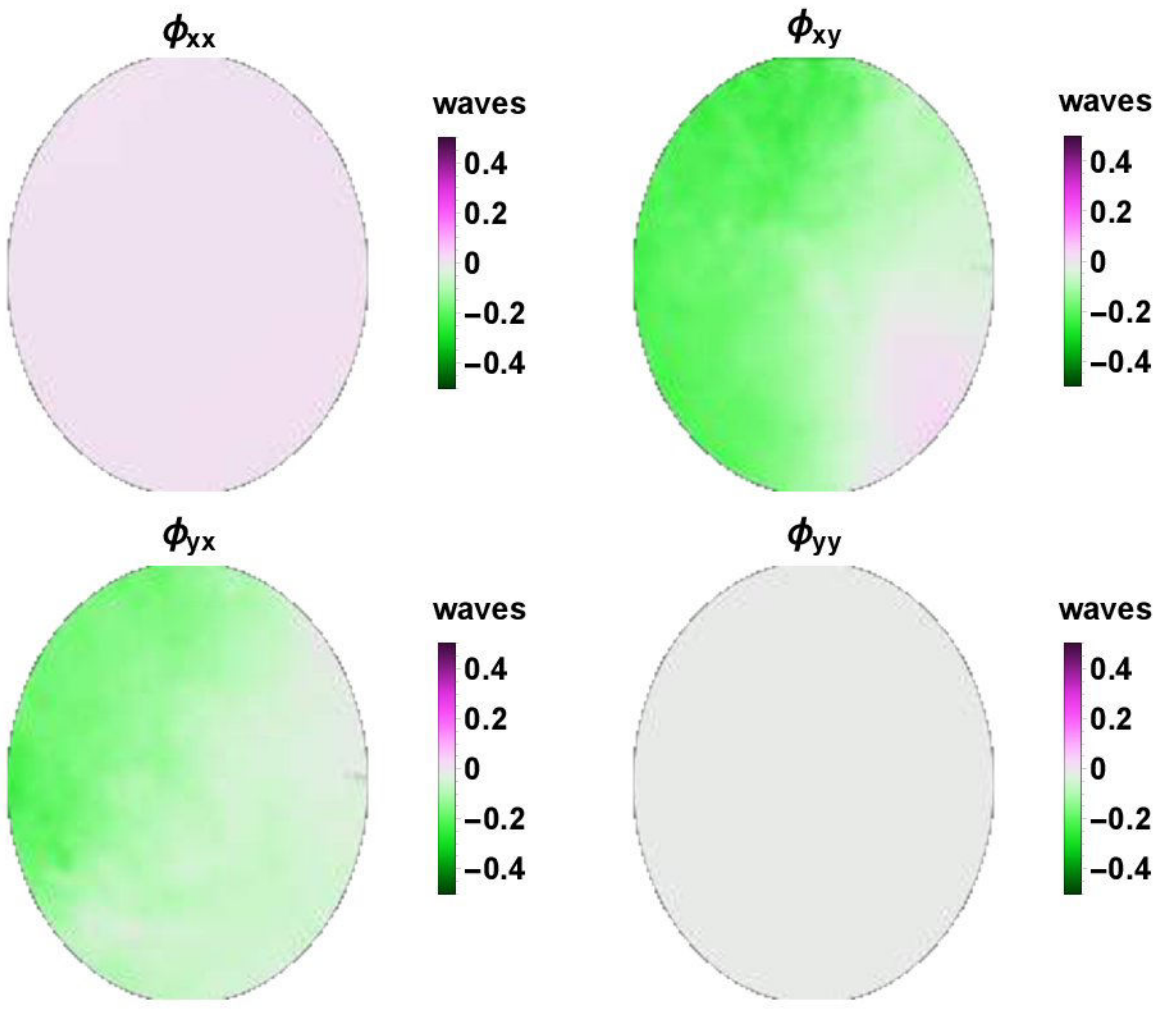

(a) 450 nm

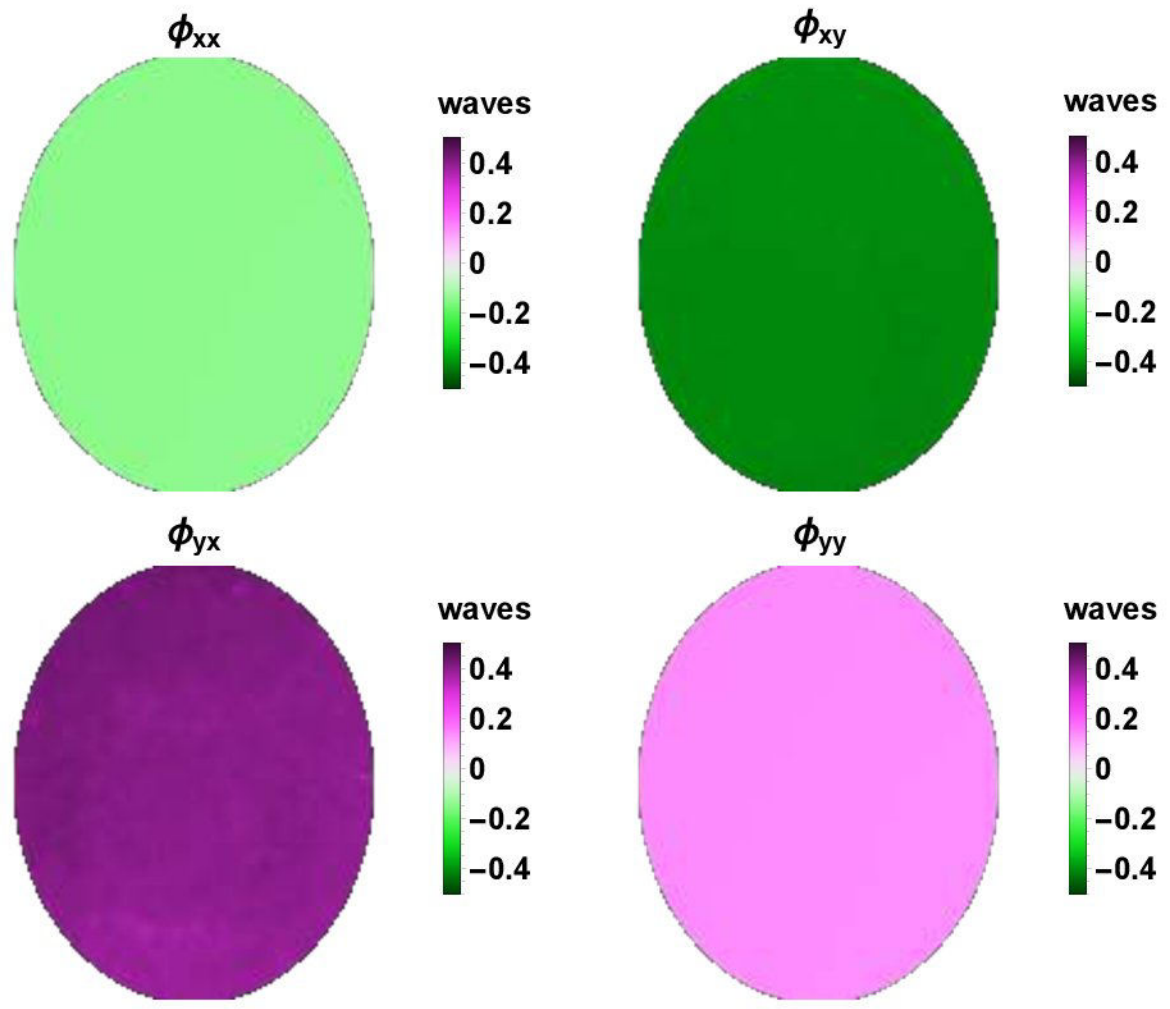

(b) 630 nm

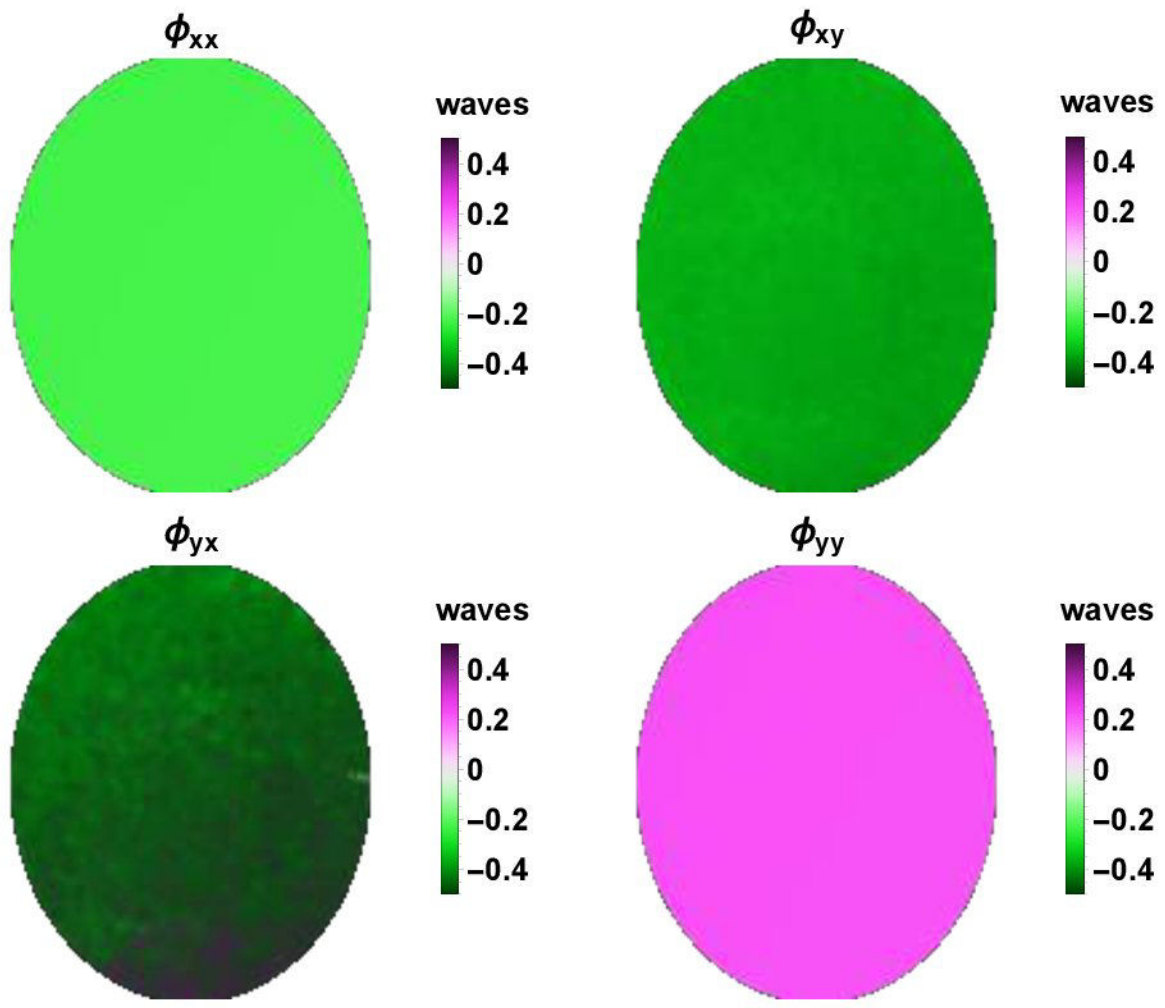

(c) 650 nm

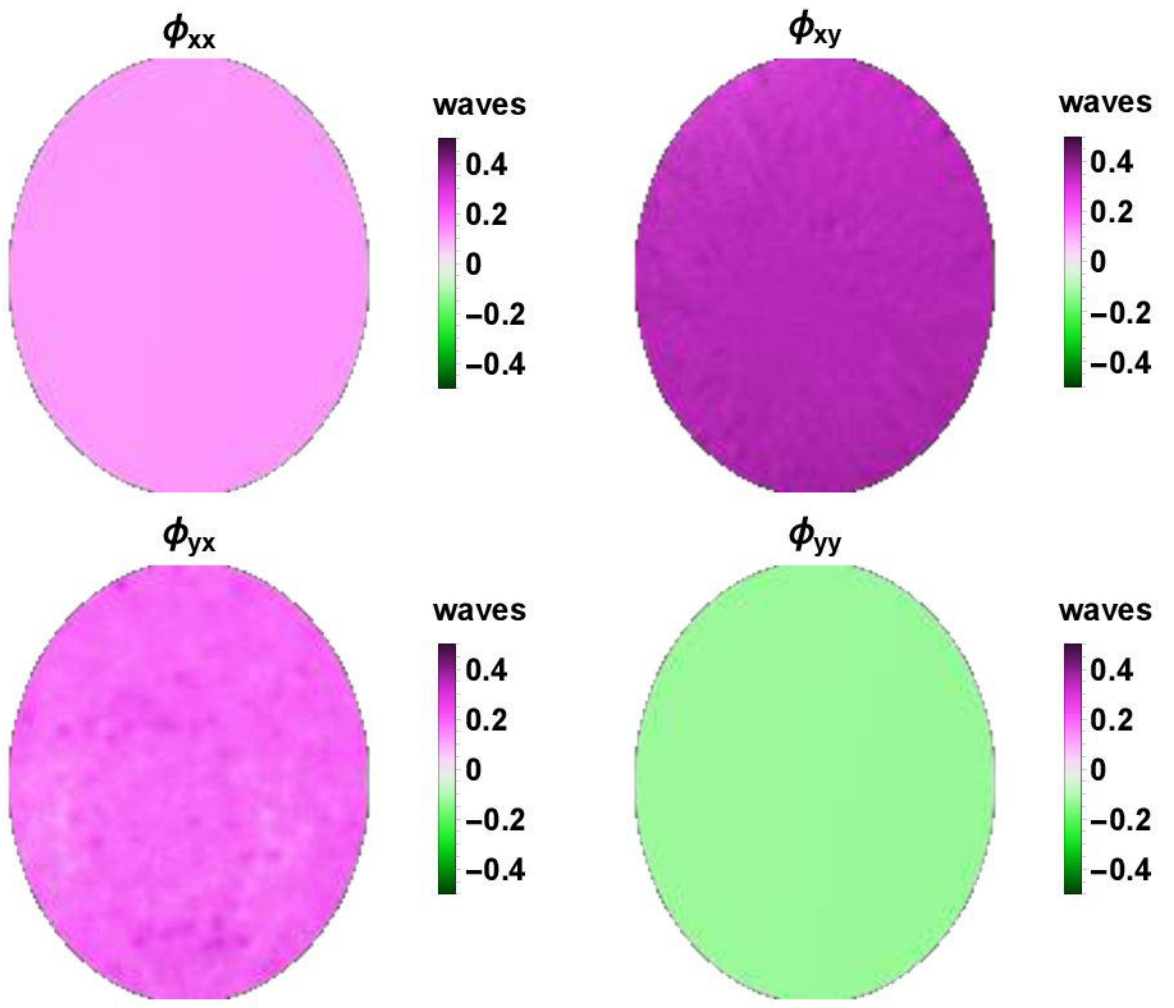

(d) 670 nm

Figure 7: The phase of the measured 2×2 Jones matrix pupils, $\phi$ of the dichroic are presented. In (a)-(d) the $\phi_{xx}$, $\phi_{xy}$, $\phi_{yx}$, $\phi_{yy}$ components are displayed for 450, 630, 650, and 670 nm.

## 4. CONCLUSION

This work has determined experimentally that dichroics introduce retardance and spatial diattenuation and thus cannot be treated as polarization invariant optical elements. Dichroics that are placed upstream of polarization sensitive optics, such as VVCs, would affect the system performance without corrective optics. A VVC is a halfwave retarder with a spatially varying fast axis. Retardance introduced upstream of the VVC will rotate the polarization state of incident light. This rotation prior to the VVC could interfere with the high contrast design of the VCC optical system. It is well know that instrumental polarization that is not corrected for is an important consideration in exoplanet imaging.[3] A dichroic introduces retardance into the system which is unavoidable due to thin film effects. The retardance introduced by the dichroic is of greater concern if incident light is polarized or partially-polarized. Unpolarized light is unaltered by a retarder. It is therefore suggested that the dichroic should not be placed immediately before or after polarization sensitive optics without proper corrections. For example, corrective waveplates could correct for unwanted retardance.[15]

A majority of optical systems are effectively insensitive to retardance and minor amounts of diattenuation introduced by a dichroic filter. However, for highly polarization sensitive systems such as VVCs, what would normally be considered negligible instrumental polarization error becomes significant. This work has shown that the requirements on the polarization properties of the dichroic need to be specified in polarization critical optical systems. As next generation coronagrahs push towards the physical limitations of performance, even small variations in diattenuation and retardance introduced by a dichroic has the potential to reduce contrast. If industry does not have the technology to fabricate a dichroic to these requirements, then NASA should establish a technology development priority and support the technology effort.

## ACKNOWLEDGMENTS

The authors thank Jeremy Parkinson of the University of Arizona's Polarization Lab and Jaren Ashcraft of the University of Arizona's Space and Astrophysics Lab (UASAL) for their help with instrument and device management.